\documentclass{PoS}

\usepackage{subcaption}
\usepackage{multirow, makecell}
\usepackage{booktabs}
\usepackage{amsmath,amssymb}

\title{MuPix and ATLASPix -- Architectures and Results}

\ShortTitle{MuPix and ATLASPix -- Architectures and Results}

\author{
\speaker{A.~Sch\"oning}$^a$,
J.~Anders$^b$,
H.~Augustin$^a$,
M.~Benoit$^c$,
N.~Berger$^d$,
S.~Dittmeier$^a$,
F.~Ehrler$^e$,
A.~Fehr$^b$,
T.~Golling$^c$,
S.~Gonzalez Sevilla$^c$,
J.~Hammerich$^a$\thanks{now at University of Liverpool, UK}~,
A.~Herkert$^a$,
L.~Huth$^a$\thanks{now at DESY, Hamburg, Germany}~,
G.~Iacobucci$^c$,
D.~Immig$^a$,
M.~Kiehn$^c$,
J.~Kr\"oger$^f$,
F.~Meier$^{ag}$,
A.~Meneses Gonzalez$^a$,
A.~Miucci$^b$,
L.O.S.~Noehte$^a$,
I.~Peric$^e$,
M. Prathapan$^e$,
T.~Rudzki$^a$,
R.~Schimassek$^e$,
D M S~Sultan$^c$,
L.~Vigani$^a$,
A.~Weber$^{a}$,
M.~Weber$^b$,
W.~Wong$^c$,
E.~Zaffaroni$^c$,
H.~Zhangh$^e$\\
\llap{$^a$} Physics Institute, Heidelberg University,  \\
Im Neuenheimer Feld 226, 69120 Heidelberg, Germany \\
\llap{$^b$} Physics Institute, University of Bern,\\
Sidlerstrasse 5, 3012 Bern, Switzerland \\
\llap{$^c$} Department  of Nuclear Physics, University of Geneva, \\
24, quai Ernest-Ansermet, CH-1211 Gen\`eve 4, Switzerland\\
\llap{$^d$} Institute for Nucelar Physics, University of Mainz, \\
Johann-Joachim-Becher-Weg 45, 55128 Mainz \\
\llap{$^e$} Institute of Data Processing and Electronics, Karlsruhe Institute of Technology,\\
P.O. Box 3640, 76021 Karlsruhe, Germany \\
\llap{$^f$} Organisation Europ\'eenne pour la Recherche Nucl\'eaire (CERN),\\
1211 Geneva 23, Switzerland \\
\llap{$^g$} Paul-Scherrer Institute, Villigen, \\
Forschungsstrasse 111, 5232 Villigen, Switzerland \\
E-mail: \email{schoning@physi.uni-heidelberg.de}
}

\abstract{High Voltage Monolithic Active Pixel Sensors (HV-MAPS) are based on
a commercial  High Voltage CMOS process and collect charge by drift
inside a reversely biased diode. HV-MAPS represent a promising technology for
future pixel tracking detectors.
Two recent developments are presented. The MuPix has a continuous readout and
is being developed for the Mu3e experiment
whereas the ATLASPix is being developed for LHC applications with a triggered readout.
Both variants have a fully
monolithic design including state machines, clock circuitries and serial
drivers. Several prototypes and design variants were characterised in the lab
and in testbeam campaigns to measure efficiencies, noise, time resolution and
radiation tolerance. 
Results from recent MuPix and ATLASPix prototypes are presented and prospects
for future improvements are discussed.}

\FullConference{The 28th International Workshop on Vertex Detectors - Vertex2019\\
		13-18 October, 2019\\
		Lopud, Croatia}

\begin{document}
\begin{figure}
\centering
\includegraphics[width=0.4\textwidth]{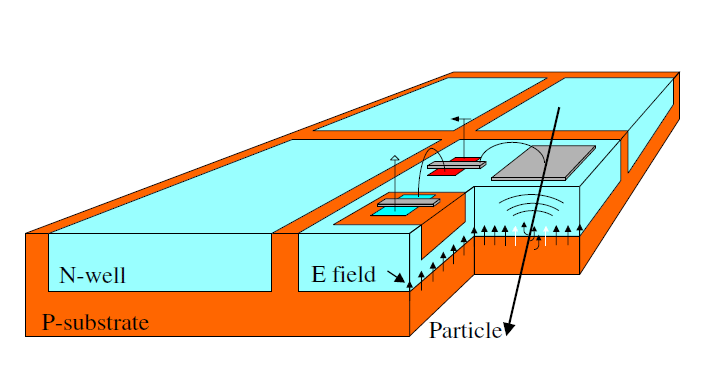}
\vspace{-0.4cm}
\caption{\textsl{Schematic drawing of an HV-MAPS \cite{peric}.}}
\label{fig:hvmaps}
\end{figure}
Future particle physics experiments have increasing demands concerning particle rates, time and spatial resolution, and radiation tolerance.
For the construction of semiconductor trackers,
monolithic active pixel sensors (MAPS) represent a very attractive alternative to standard hybrid designs where the sensors and readout chips are on different dies. Hybrid designs require interconnects like wires or bump bonds which are a limiting factor for the maximum achievable spatial resolution and the scalability when it comes to instrumenting large areas.
Particularly interesting are depleted MAPS which collect ionisation charge mainly via drift, provide time resolutions 
of a few nanoseconds and are radiation tolerant.

\section{High Voltage Monolithic Active Pixel Sensors (HV-MAPS)}

\begin{figure}[t!]
\centering
\includegraphics[width=0.4\textwidth]{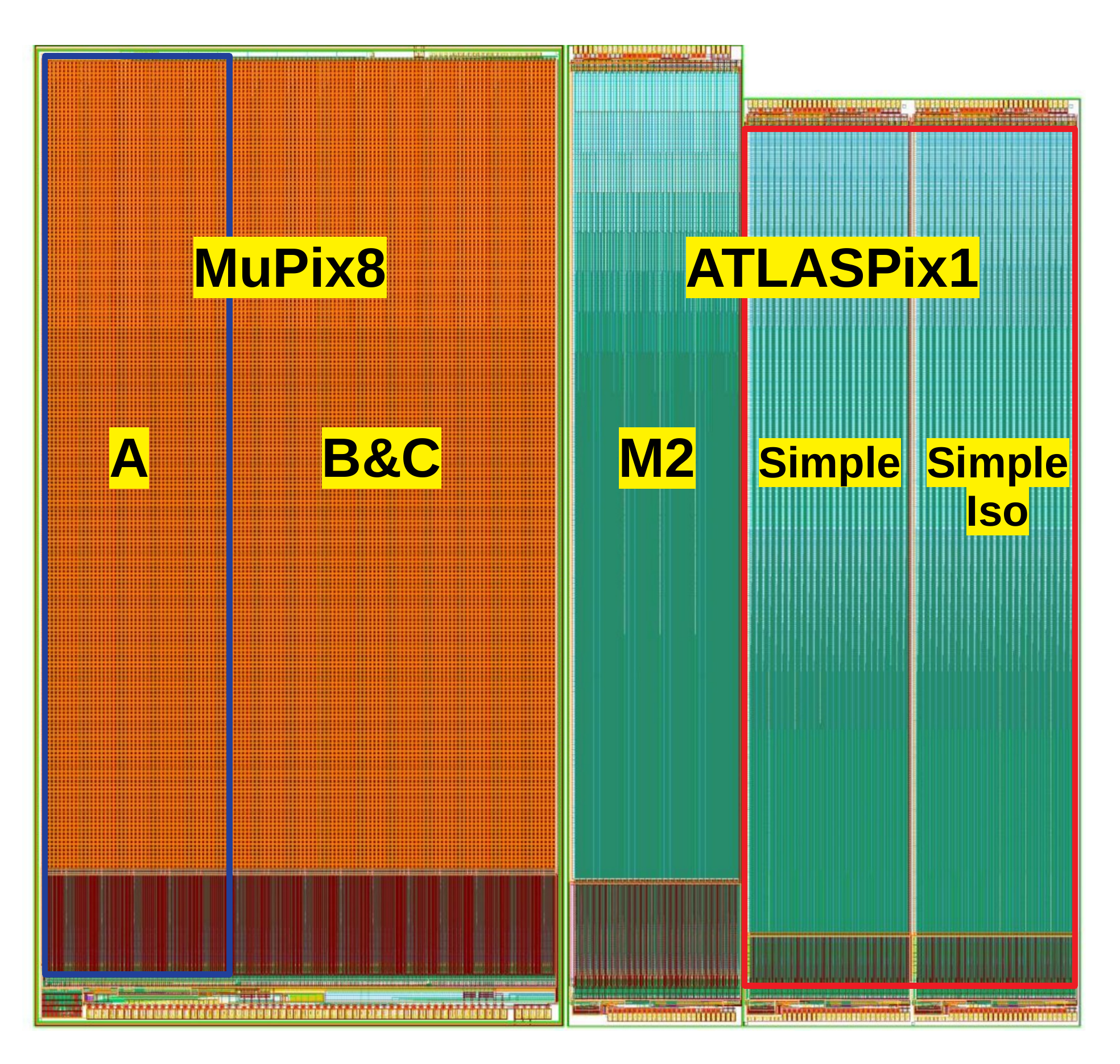}
\vspace{-0.1cm}
\caption{\textsl{CAD views of MuPix8 and ATLASPix1 prototypes as produced on the same reticle.}}
\label{fig:mupix_atlaspix}
\end{figure}

In contrast to ordinary MAPS designs where the charge collecting diode and the pixel electronics are spatially separated,
HV-MAPS exploit HV-CMOS processes and implement the pixel electronics inside a deep n-well~\cite{peric}.
HV-MAPS collect ionisation charge mainly via drift, see figure~\ref{fig:hvmaps}, and provide time resolutions  of a few nanoseconds in contrast to standard MAPS
which collect ionisation charge mainly by diffusion. The depletion depth grows with the bias voltage and the substrate resistivity.
Breakdown voltages of up to 200\,V have been realised, depending on the substrate resistivity and the electrical field in the pixels (e.g. guard ring design). 
The fast collection of the signal charges and the low noise allow for the implementation of continuous readout schemes where zero suppressed hits are only temporarily buffered and sent out using fast Gbit/s links.

 HV-MAPS have been prototyped for several experiments (Mu3e, ATLAS, CLIC) with different readout architectures. HV-MAPS are also considered for other
particle and nuclear physics experiments.
All designs have in common that the readout circuitry is spatially separated from the active pixel matrix.
They provide fast time stamping and charge measurement using time-over-threshold (ToT), thus allowing for
timewalk correction.
Due to the small size of the active depletion zone of about 
15-50~$\mu$m for substrate resistivity of 20-200\,$\Omega$cm, 
HV-MAPS detectors can be thinned down to 50-100\,$\mu$m.
This makes the HV-MAPS concept interesting for low energy experiments like Mu3e, where charged particle tracking is limited by multiple scattering.
For the silicon pixel detector of the Mu3e experiment~\cite{RP_Mu3e}, aiming to search for the decay $\mu^+ \rightarrow e^+e^+e^-$ with unprecedented sensitivity,
the MuPix sensor is being developed.

The ATLASPix development, on the other hand,  was originally intended to provide for the new ATLAS inner tracker system (ITk) at High Luminosity LHC
an alternative sensor technology for the outermost pixel layer at a radius of~29\,cm.
Several HV-MAPS prototypes, submitted in a common engineering run in the year 2017, are shown in figure~\ref{fig:mupix_atlaspix}.
In the following the focus is put on results obtained from MuPix8 (submatrix A) and ATLASPix1 (version \textit{simple}) characterisation studies.
A main difference between the sensors is the way how signals are transferred from the active matrix to the periphery.
All MuPix sensors amplify the collected charges in the pixel cell and send the amplified signal to the periphery where the comparators are located.
The submatrix~A of MuPix8 implements a source follower whereas submatrices B\&C implement a current driver.
All ATLASPix sensors have the comparators in the pixel cell and send the discriminated output to the periphery where the readout logic is implemented.
The readout logic and state machine of MuPix8 and ATLASPix1\_Simple are very similar and provide data streaming (continuous readout).
ATLASPix1\_M2 implements trigger buffers and is not discussed here.
ATLASPix1 was produced in different flavours: In ATLASPix1\_Simple the discriminator is realised in NMOS-logic whereas CMOS-logic is used
in the ATLASPix1\_Simple\_Iso.
An isolating deep p-well is implemented here to better shield the charge sensitive diode from the p-type transistors.

All prototypes discussed here have radiation tolerant designs, including the use of enclosed layout transistors where necessary.
Signal collection in the diode is also radiation tolerant, since charge is collected over short drift lengths and not strongly affected by bulk damage.

\begin{table*}[b!]
        \centering
        \small
        \begin{center}
                \begin{tabular}{l|cccc}
                                             &  MuPix8            & MuPix10          & ATLASPix1\_Simple   &  ATLASPix3       \\  \midrule
  Process                                    & AH18 (AMS)~\cite{ref:ams}         & H18 (TSI)~\cite{ref:tsi}        & AH18 (AMS)~\cite{ref:ams}         & H18 (TSI)~\cite{ref:tsi}  \\
  Sensor size [mm$^2$]                       &  $10.7\times 19.5$ & $20.2\times23.0$ & $3.4  \times 18.4$ & $20.2 \times 21.0$  \\
  Pixel matrix                              &  $128 \times 200$  & $250 \times 256$ & $25 \times 400$    & $132 \times 372$   \\
  Pixel size [$\mu$m$^2$]                    & $81\times80$       & $80\times80$     & $130\times 40$     & $150\times 50$    \\
  Active area [mm$^2$]                       &  $10.3\times 16.0$ & $20.1\times20.0$ & $3.25 \times 16.0$ & $19.8 \times 18.6$   \\
  \#Bits timestamp \& ToT                   &      10+6            &      11+5          & 10+6                 & 10+7             \\ 
  Pixel tune DACs  [bits]                   &      3              &      3           & 3                  & 3               \\ 
  Bandwidth [Gbit/s]                        & $\leq 1.6$ ~(3$\times$)      & $\leq$ 1.6  ~(3$\times$)  &  $\leq$ 1.6          & $1.28$    \\ 
  Readout architecture                       &  continuous        & continuous       & continuous          & trigger (cont.)   \\
  Status (December 2019)               & tested             & in production    & tested             & under test     \\ \bottomrule
                \end{tabular}
        \caption{\textsl{Main specification parameters of the HV-MAPS sensors discussed in this talk.}}
        \label{tab:HVMAPS_Overview}
        \end{center}
\end{table*}

The main specification parameters of the HV-MAPS sensors discussed in this talk are shown in table~\ref{tab:HVMAPS_Overview}.
The MuPix8 and ATLASPix1\_Simple sensors were produced in the AH18 (180~nm) HV-CMOS process by ams~AG (AMS)~\cite{ref:ams} and have been fully characterised in the lab and in several test beams at CERN, DESY, FERMILAB and PSI.

\section{MuPix Design and Results}
\label{sec:mupix}
\begin{figure}
\centering
\includegraphics[width=0.67\textwidth]{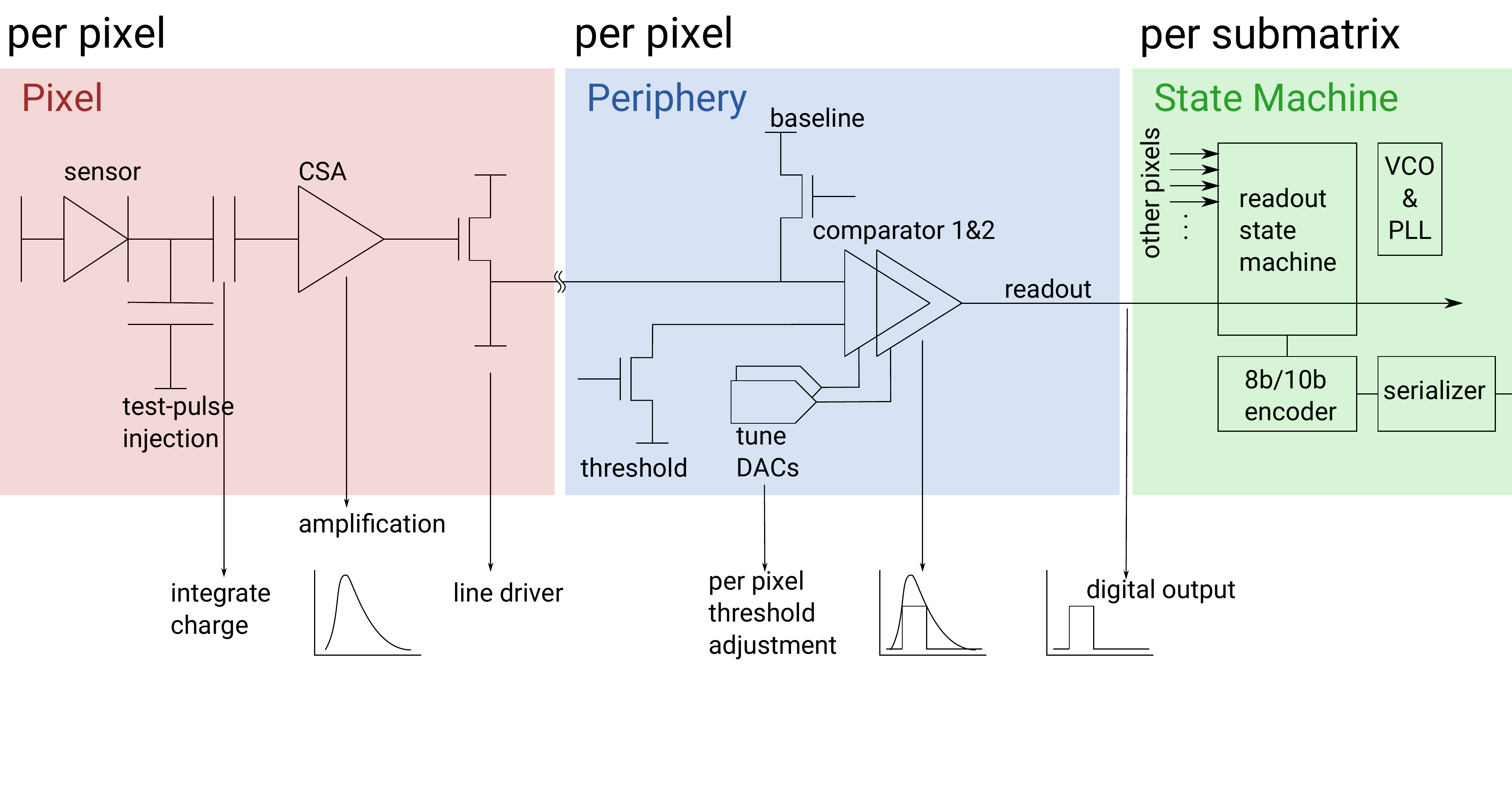}
\vspace{-0.8cm}
\caption{\textsl{Schematics of the MuPix8 readout.}}
\label{fig:mupix8_diagram}
\end{figure}

\begin{figure}[b!]
\centering
\includegraphics[width=0.6\textwidth]{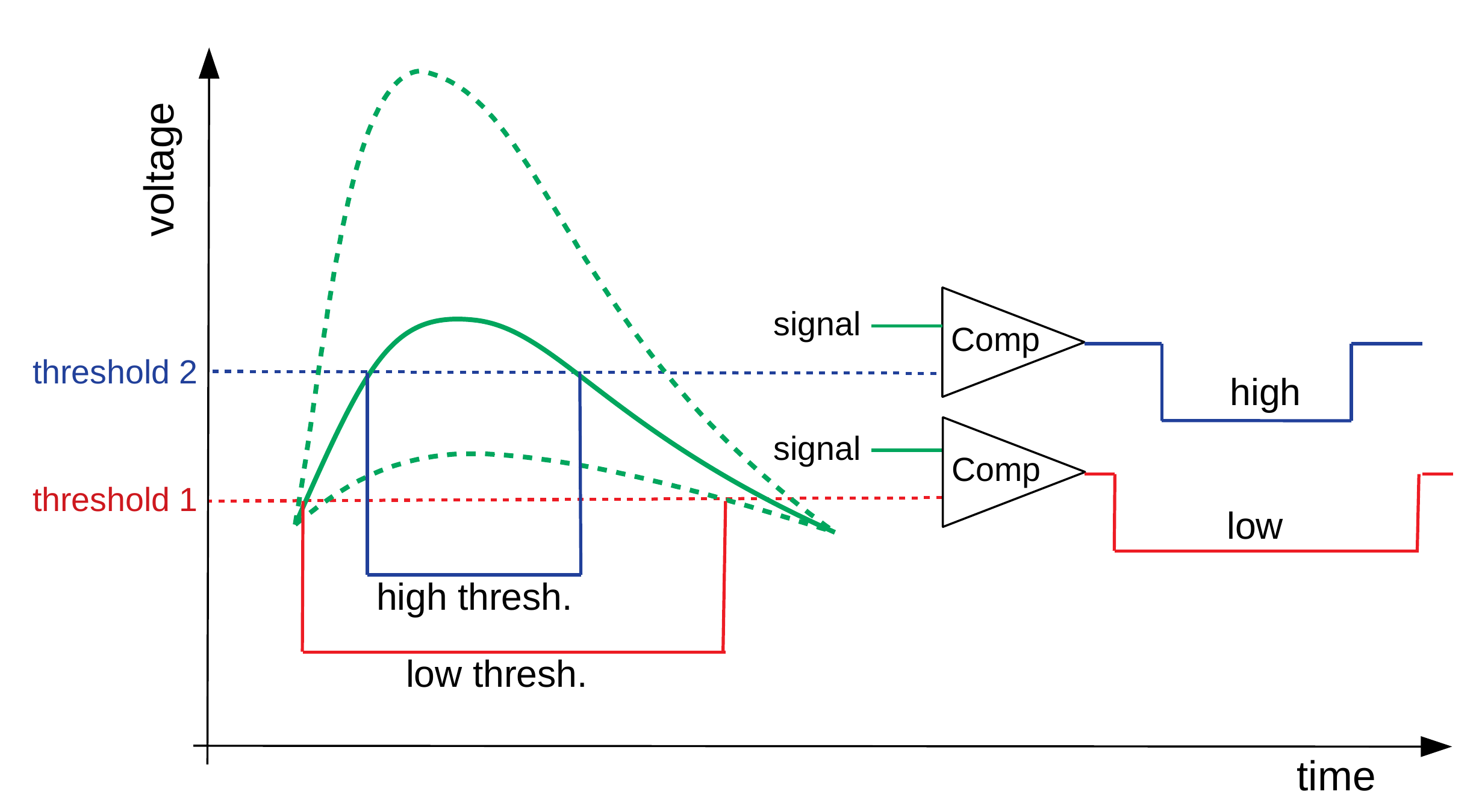}
\vspace{-0.1cm}
\caption{\textsl{The 2-threshold concept of MuPix8.}}
\label{fig:mupix8_comparators}
\end{figure}

All MuPix prototypes implement the amplifier in
the active pixel cell and the comparator in the periphery, both interconnected by an
analogue readout line.
The periphery of all recent prototypes contains in addition: 
readout buffers, the state machine, PLL and VCO, 8/10 bit encoders, and serializers providing data output rates of up to 1.6\,Gbit/s per link, see figure~\ref{fig:mupix8_diagram}.

The MuPix8 features two comparators per pixel which allow for different operation modes for timewalk correction and mitigation.
The 2-threshold method mitigates timewalk by using a very low threshold for time stamping and a higher threshold for hit validation, see figure~\ref{fig:mupix8_comparators}.
This allows the sensor to operate with low noise (high threshold) and low jitter (low threshold). 
In addition, timewalk can be corrected offline by exploiting the 6-bit ToT information.
Results of time resolution measurements are discussed in section~\ref{sec_time}.

\begin{figure}
\centering
\begin{subfigure}[l]{0.45\textwidth}
\includegraphics[width=0.95\textwidth]{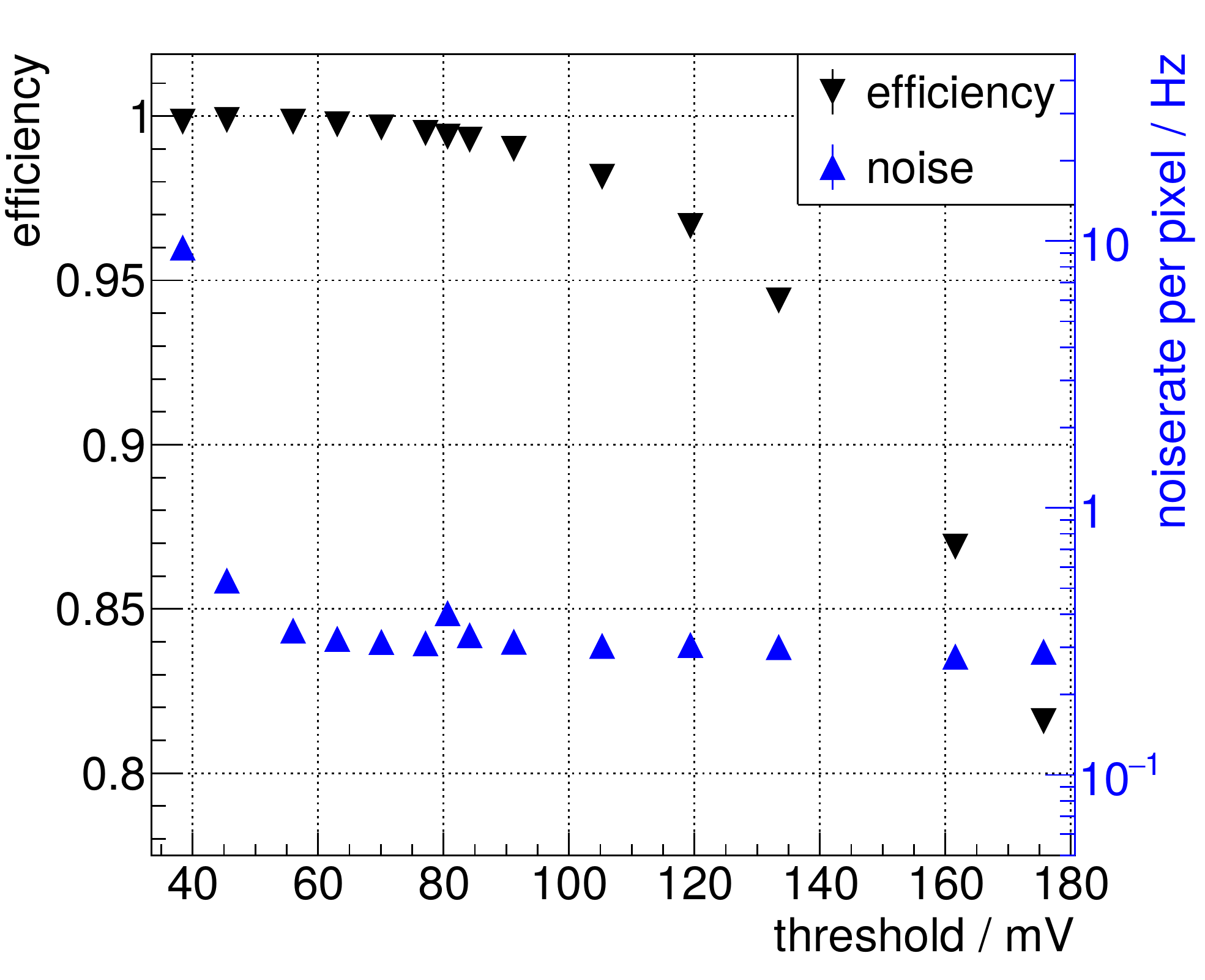}
\subcaption{\textsl{Efficiency and noise}}
\label{fig:mupix8_eff_noise}
\end{subfigure}
\begin{subfigure}[r]{0.54\textwidth}
\includegraphics[width=0.99\textwidth]{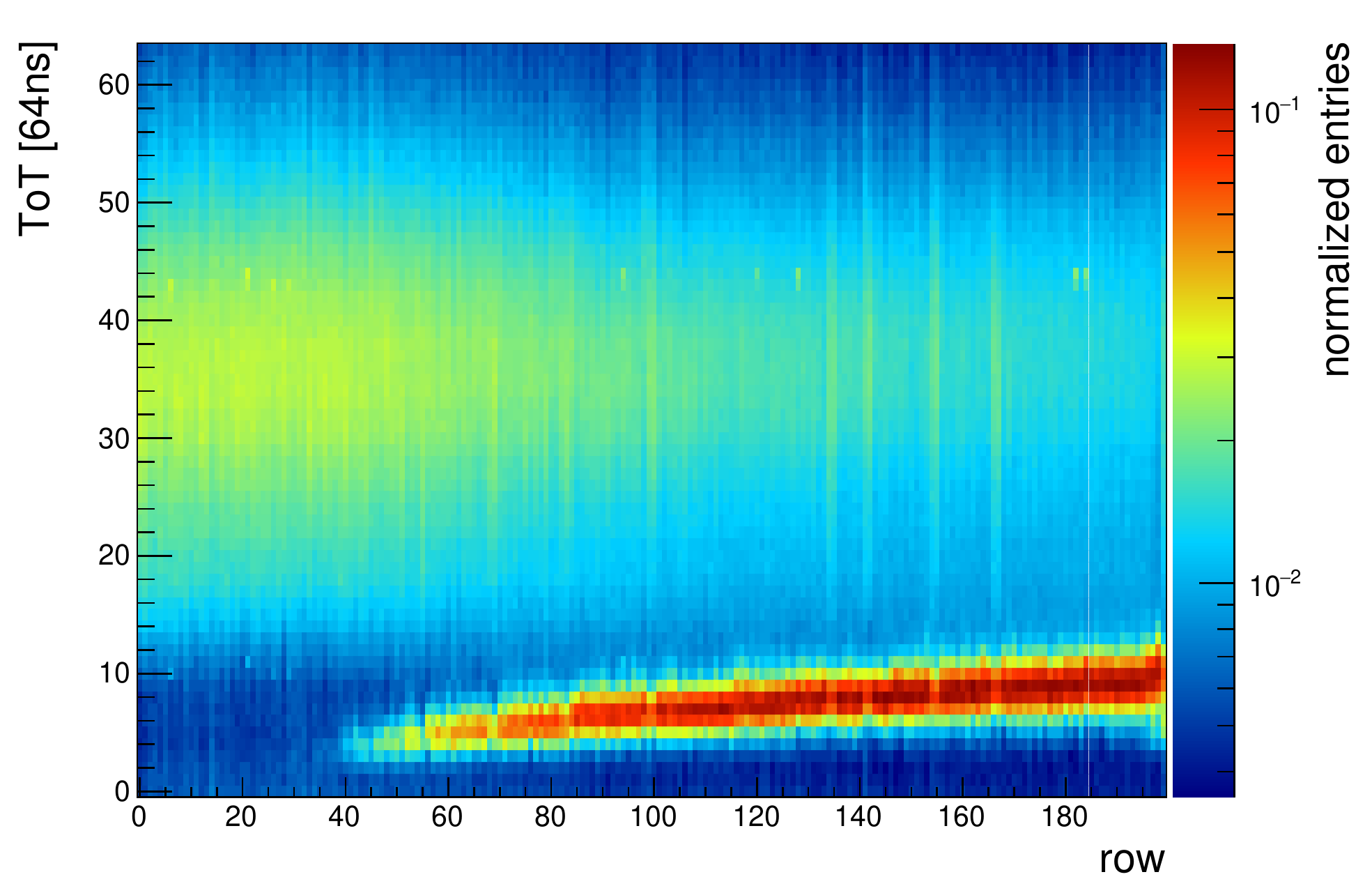}
\subcaption{\textsl{ToT versus row number}}
\label{fig:ToT_rows}
\end{subfigure}
\vspace{-0.1cm}
\caption{\textsl{Left: Efficiency and noise of MuPix8 with a substrate resistivity of 80\,$\Omega$cm as function of the threshold voltage. A threshold of 100\,mV corresponds to about 1350e$^-$. Right: ToT distribution (normalised) as function of the pixel row number. Plots taken from~\cite{hammerich,huth}.}}

\end{figure}

Efficiency and noise have been measured in test beams as a function of the comparator threshold. Figure~\ref{fig:mupix8_eff_noise} shows the results obtained for a MuPix8 sensor produced with a substrate resistivity of 80\,$\Omega$cm. A plateau of high efficiency is reached for thresholds below 80~mV. For higher thresholds, inefficiencies arise from charge sharing effects at pixel edges and corners.
For thresholds above 50\,mV the noise rate is very low and significantly below 1\,Hz per pixel. For thresholds below 50\,mV the noise rate significantly increases. In the range 50-80\,mV a stable operation of MuPix8 is possible, providing high efficiency above 99\% at low noise.

A disadvantage of the MuPix8 design
are the long interconnects between active pixel matrix and periphery of up to 2\,cm length which cause large interline capacitances.
The resulting cross talk crucially depends on the spacing of the readout lines.
The MuPix8 sensor has six metal layers and uses two for signal lines.
Measurements have shown significant cross talk which slightly compromises
the hit detection efficiency for pixels with long routings. The cross talk between the interconnects can be seen in figure~\ref{fig:ToT_rows} which shows the ToT distribution as function of the row number of the pixel. For small row numbers (short interconnects) a signal peak is visible for ToT$=35$.
For row numbers $\ge 40$ a second peak appears originating from cross talk.

\section{ATLASPix}

An alternative readout architecture which implements the comparator in the 
pixel cell has been developed for ATLAS at LHC.
This architecture is practically immune against cross talk from interline capacitances
and has the advantage that the signal is discriminated just after amplification, and
thus not deteriorated by driving the analogue signal over long lines like in the MuPix design.
A disadvantage is the larger pixel capacitance from the comparator logic inside the pixel cell.

\begin{figure}
\centering
\begin{subfigure}[l]{0.4\textwidth}
\includegraphics[width=\textwidth]{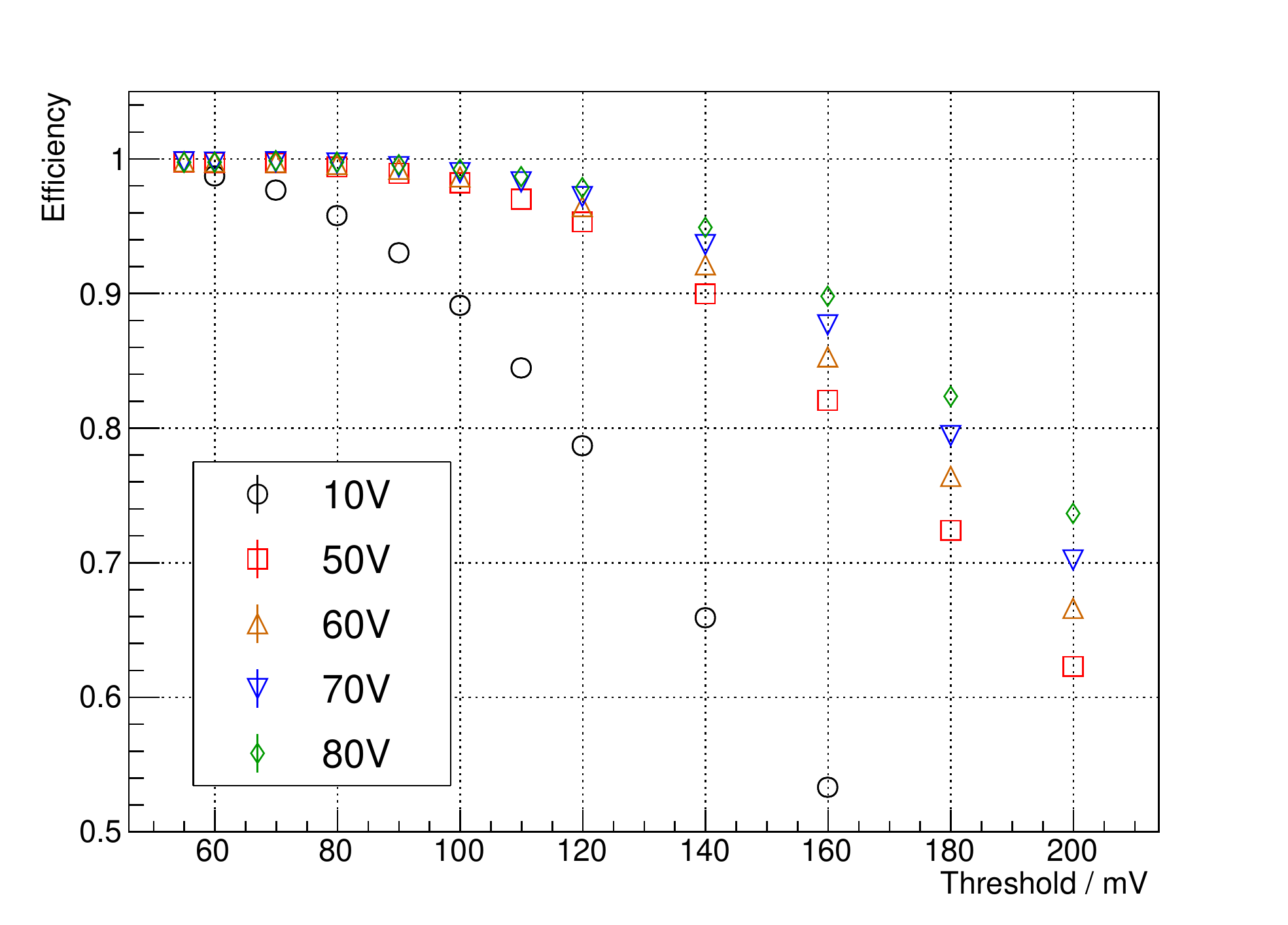}
\subcaption{\textsl{substrate resistivity 80\,$\Omega$cm}}
\end{subfigure}
\begin{subfigure}[r]{0.4\textwidth}
\includegraphics[width=\textwidth]{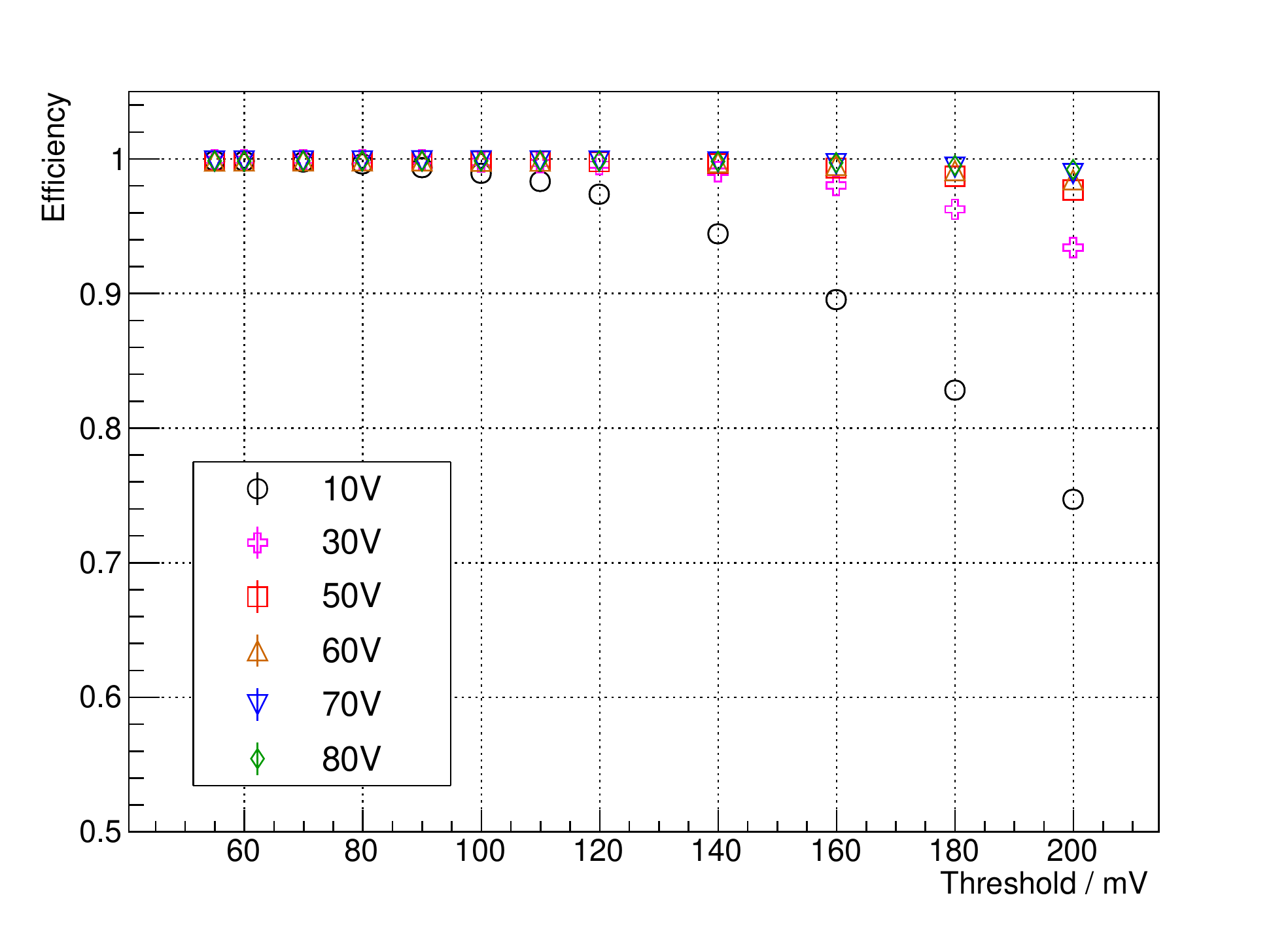}
\subcaption{\textsl{substrate resistivity 200\,$\Omega$cm}}
\end{subfigure}
\caption{\textsl{Single hit efficiencies of ATLASPix1\_Simple for two different substrate resistivities as function of the discriminator threshold and for different bias voltages. The measurement were performed with pions of 300\,MeV/c at the PSI testbeam. Plots taken from \cite{herkert}. }}
\label{fig:atlas_eff}
\end{figure}

First results from ATLASPix1 characterisation studies at CERN and FERMILAB have been reported in~\cite{Kiehn:2019wpe,Peric:2019hmv}.
More results were obtained at recent test beam campaigns at DESY and PSI. Single hit efficiencies have been measured and are shown in figure~\ref{fig:atlas_eff}
for samples with substrate resistivities of 80\,$\Omega$cm and 200\,$\Omega$cm as a function of the discriminator threshold and different bias voltages.
For bias voltages above 50\,V a high efficiency plateau is visible for a threshold range of 50-100 (50-200) mV for substrate resistivities of  80 (200)\,$\Omega$cm.
For bias voltages of 70-80\,V single hit efficiencies exceeding $99.7\,\%$ are reached with practically negligible noise rates.

\section{Time Resolution}
\label{sec_time}
The time resolutions of the MuPix8 and ATLASPix1 sensors are measured with a $^{90}$Sr source.
For MuPix8, the measurement is performed for different operation modes of the two comparators (see section~\ref{sec:mupix}).
Operating the sensor using a common comparator threshold of 50\,mV, corresponding to about 650\,e$^-$, a time resolution of 7.6\,ns (10.5\,ns)
is measured with (w/o) timewalk correction.
Using the 2-threshold method with a lower threshold at 35\,mV, the time resolution improves to 6.7\,ns (8.8\,ns), respectively.
The two-threshold method therefore helps to improve the time resolution, even if an offline timewalk correction is applied.

\begin{figure}[b]
\centering
\begin{subfigure}[l]{0.34\textwidth}
\includegraphics[width=1.0\textwidth]{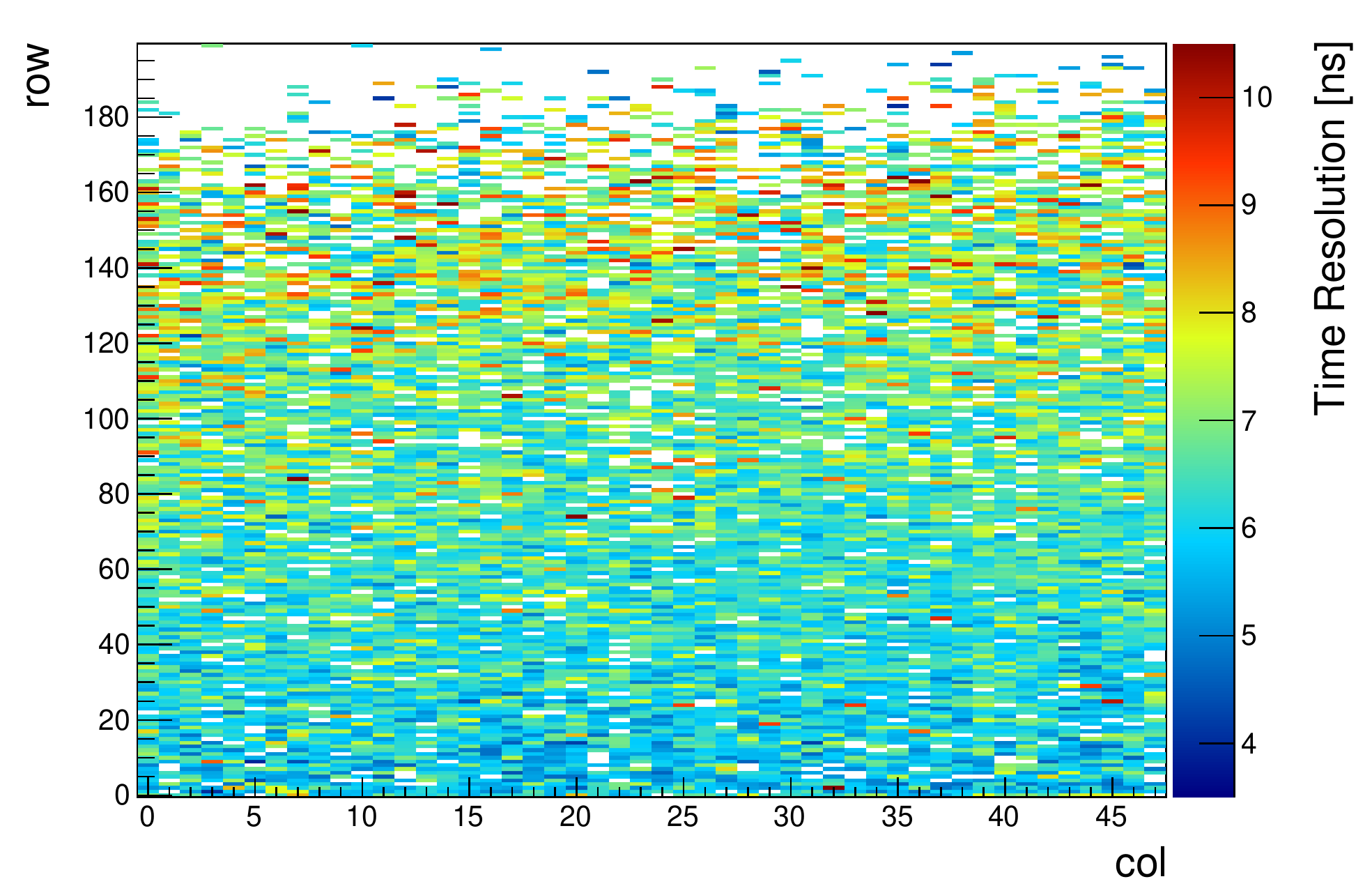}
\subcaption{\textsl{MuPix8 submatrix A}}
\end{subfigure}
\begin{subfigure}[m]{0.32\textwidth}
\includegraphics[width=\textwidth]{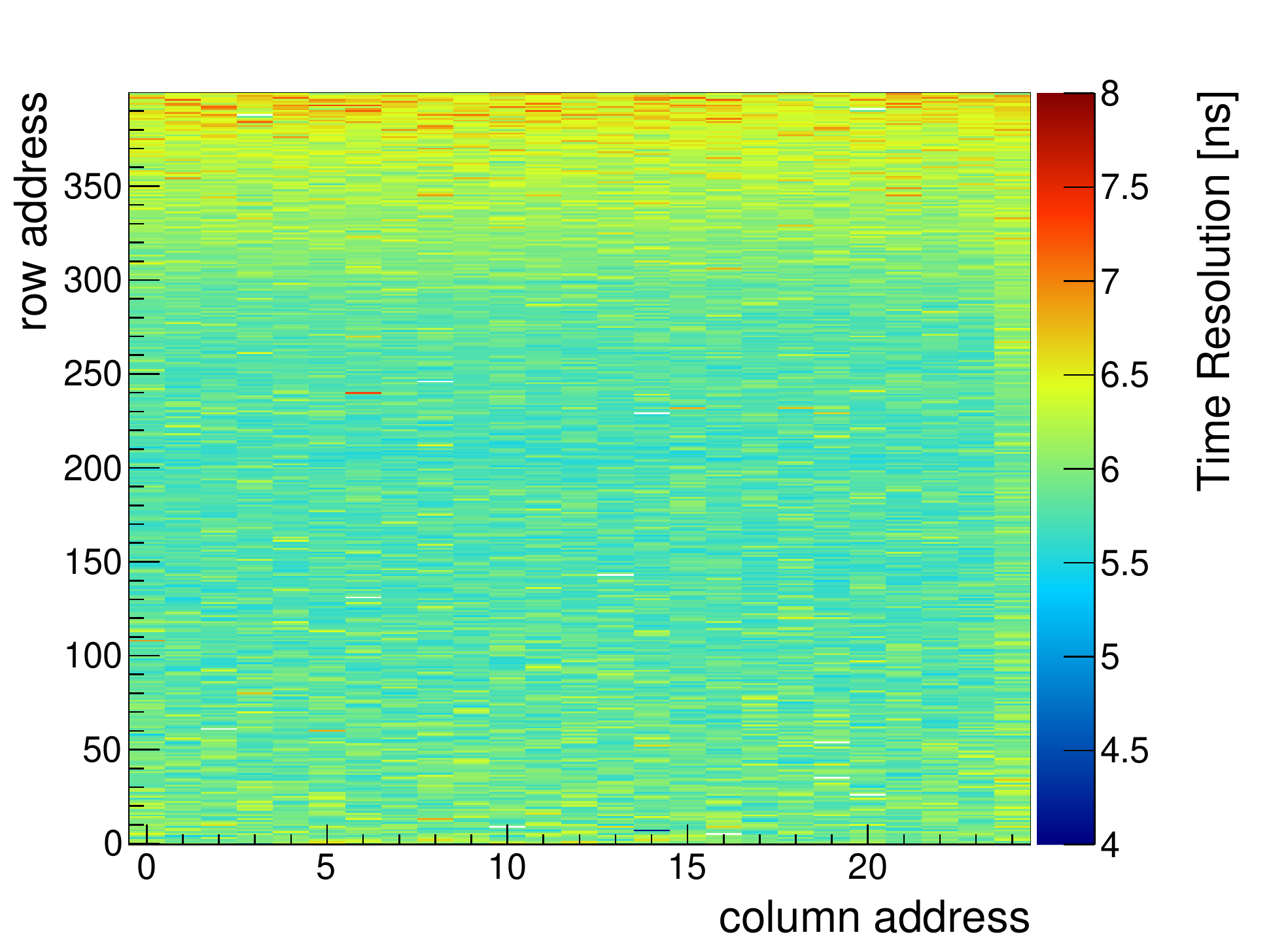}
\subcaption{\textsl{ATLASPix1}}
\end{subfigure}
\begin{subfigure}[r]{0.32\textwidth}
\includegraphics[width=\textwidth]{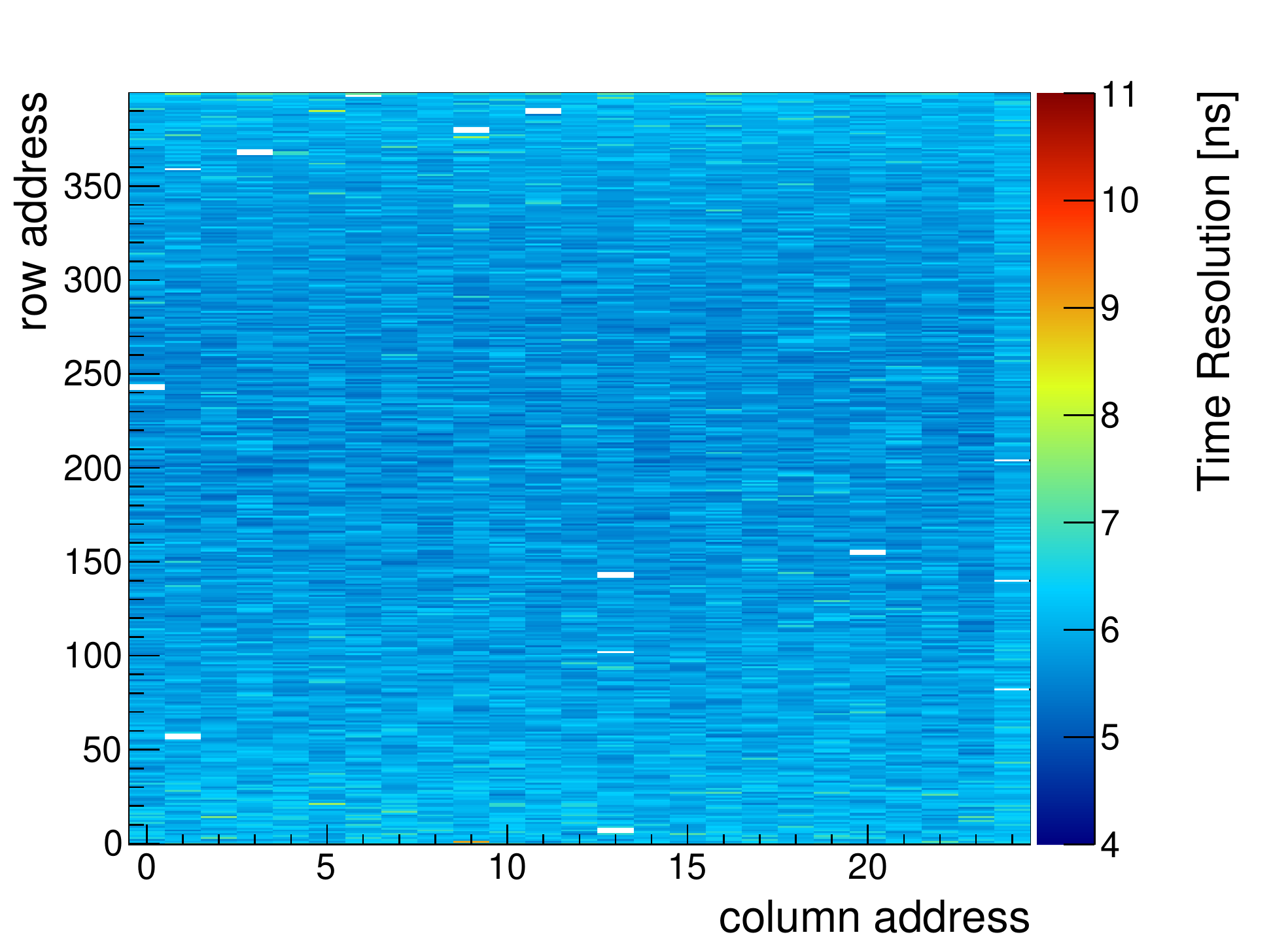}
\subcaption{\textsl{ATLASPix1\_Iso}}
\end{subfigure}
\vspace{-0.1cm}
\caption{\textsl{Maps of the time resolution of the full pixel matrix after timewalk correction as measured by a $^{90}$Sr source with a reference timing counter. Plots from \cite{hammerich,immig}.}}
\label{fig:tres_map_histo}
\end{figure}

For MuPix8, the map of time resolutions as function of the pixel region is shown in figure~\ref{fig:tres_map_histo}a for a common comparator threshold of 50\,mV.
A clear row dependence is visible: pixels at low row numbers have a better time resolution than those at large row numbers.
This suggests that the large interconnects negatively affect the analogue pulses and the time resolution.
However, an insufficient power distribution over the matrix can also not be excluded, given that power supply and bias signals originate from the periphery near the bottom row of the pixel matrix.
Similar time resolution maps are shown in figures~\ref{fig:tres_map_histo}b and~c for ATLASPix1\_Simple and ATLASPix1\_Simple\_Iso, respectively.
These distributions look much more homogeneous and for the latter there is no loss of resolution with higher row numbers, which are further from the periphery.

\begin{table}
  \small
\center
\begin{tabular}{l|c|c|c}
                           &  & \multicolumn{2}{c}{ATLASPix1} \\
                           & MuPix8       & Simple & ISO p-well \\ \midrule
Timestamp sampling [ns]  & 8.0    & 16.0     & 16.0 \\ \midrule
Time resolution  ($\sigma_\textrm{fit}$)      &        &        &  \\
-~w/o TW correction [ns]  & 8.8    & 8.1    & 6.8 \\
-~with TW correction [ns] & 6.7    & 5.9    & 5.8 \\
-~internal resolution [ns] & 6.3    & 3.7    & 3.6 \\
\end{tabular}   
\caption{\textsl{Time resolution  of the three different sensor types before and after timewalk (TW) correction using ToT information. The internal time resolution is defined by the TW-corrected resolution minus the quadratically subtracted sampling resolution given by the time stamping frequency.}}
\label{table:time_resolutions}
\end{table}

A quantitative comparison of the time resolution obtained by MuPix8 with the 2-threshold method and both ATLASPix1\_Simple versions which implement only one comparator threshold is given in table~\ref{table:time_resolutions}. All three sensors show very similar time resolutions after timewalk correction. However, integrating a different clock division scheme, the MuPix8 has a two times higher sampling frequency than both ATLASPix1 versions. Therefore, the time resolutions measured for ATLASPix1 include a significant component from the sampling.
By correcting the sampling uncertainty an internal time resolution is calculated. This time resolution is  $\sim$3.7\,ns for both ATLASPix1\_simple versions and significantly lower than the 6.3\,ns obtained for MuPix8.

\section{Irradiation Studies}
\label{sec:irrad}
\begin{figure}
\centering
\includegraphics[width=0.6\textwidth]{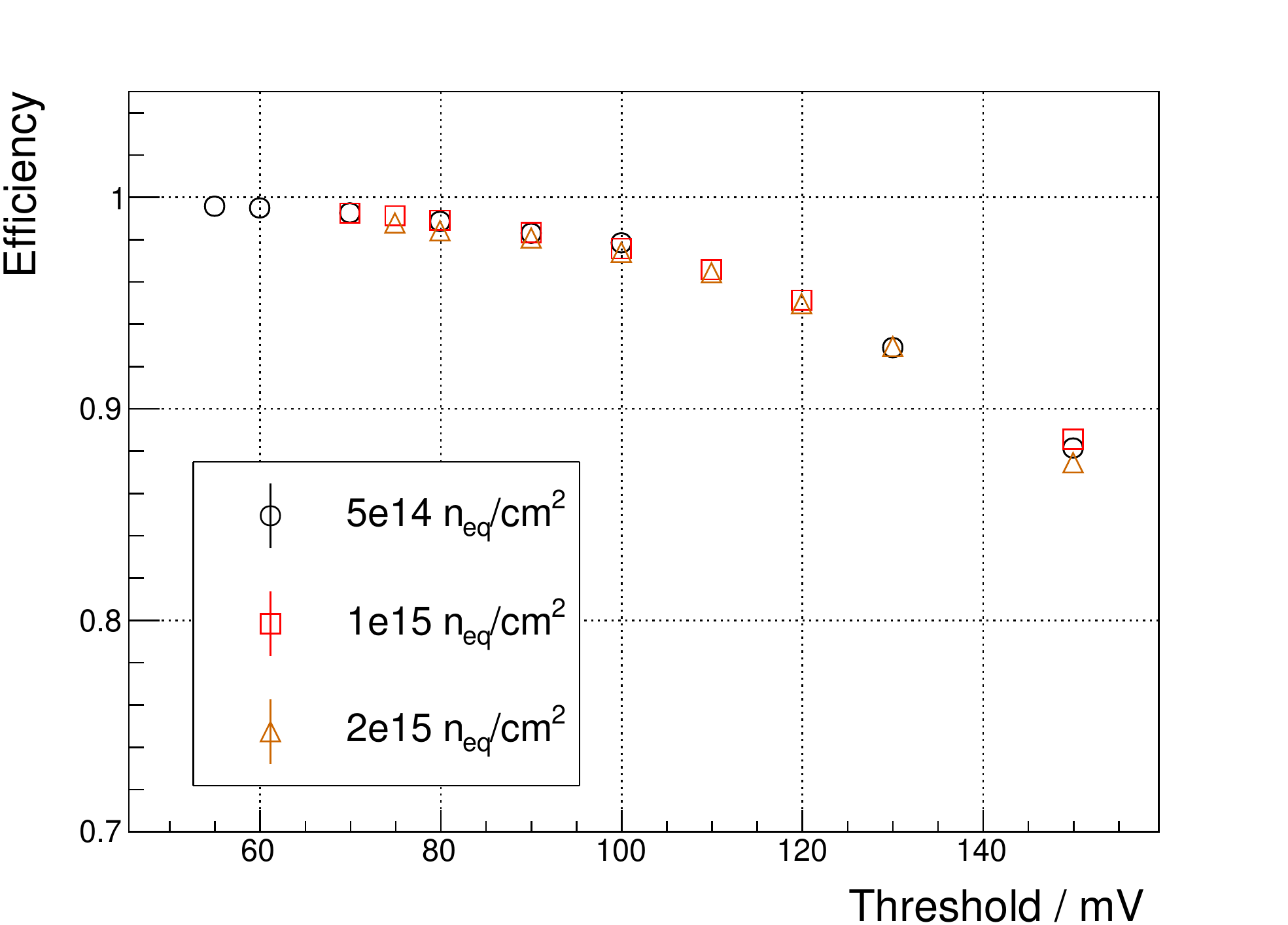}
\vspace{-0.1cm}
\caption{\textsl{Single hit efficiency for neutron irradiated ATLASPix1\_Simple with a substrate resistivity of 80\,$\Omega$cm as function of the comparator threshold for fluences of $5 \cdot 10^{14}$, $1 \cdot 10^{15}$, and $2 \cdot 10^{15}$\,n$_\textrm{eq}$/cm$^2$. The sensors are thinned to 62\,$\mu$m and a bias voltage of 60\,V is applied. All sensors are untuned and had a temperature of about 5$^\circ$C. Plot taken from~\cite{herkert}.}}
\label{fig:eff_irrad_neutron}
\end{figure}

\begin{table}[b]
  \small
\center
\begin{tabular}{l|c|r|r|r}
\multirowcell{2}{Resistivity /\\$\Omega$cm}     & \multirowcell{2}{Fluence / \\$10^{14}$ n$_\textrm{eq}$/cm$^2$}          & 
$-60$\,V              & $-70$\,V        & $-80/85$\,V           \\ \cline{3-5}
        &                                                                       & \multicolumn{3}{c}{\rule{0pt}{2.5ex}Efficiency / \%}  \\ \midrule
\multirowcell{4}{80}
        &  ~$1$~ ~(n)                   & 96.3          & 97.5          & 98.3                  \\
        &  ~$5$ ~~(n)                   & 99.5        & -                     & -                             \\
        &  $10$~ ~(n)                   & 99.3          & -                     & 99.5                  \\
        &  $20$ ~~(n)                   & 98.5          & 98.4          & 98.6                  \\ \midrule
\multirowcell{4}{200}
        & ~$0$~                         & -         &  -        &   $\ge$99.7               \\
        & ~$5$ ~~(n)                    & 99.2          & -                     & -                             \\
        &  $10$~ ~(n)                   & 98.8          & -                     & $^\dagger$99.5                             \\
        &  $20$ ~~(n)                   & 96.5          & -                     & 98.7                      \\ \midrule
        & ~$1$~ ~(p)                 & $^\dagger$99.7          &          &                  \\      
        & ~$5$ ~~(p)                 & 99.6          & 99.7          & 99.9                  \\      
\end{tabular}   
\caption{\textsl{Maximum efficiencies of proton (p) and neutron (n) irradiated ATLASPix1\_Simple sensors at an average noise rate per pixel of $\lesssim$40\,Hz.
The sensors with a wafer resistivity of $80\ \Omega$cm are thinned to $62\,\mu$m, the $200\,\Omega$cm to $100\, \mu$m, and the sensors where the efficiency is labeled with $^\dagger$ have not been thinned ($725\, \mu$m). Modified table from~\cite{herkert}.}}
\label{table:irrad_eff}
\end{table}

The radiation hardness of the AMS H18 process has been investigated in previous works~\cite{Benoit:2016vup,Hiti:2019chp} including an electrical characterisation of irradiated ATLASPix1 prototypes~\cite{Sultan:2019jhg}.
Single hit efficiencies for a fully monolithic HV-MAPS\footnote{We define a pixel sensor as being fully monolithic if the full readout architecture including VCO, PLL and state machine is implemented.} irradiated with protons and neutrons were reported first for the MuPix7 predecessor~\cite{ref:irrad_hvmaps} which has no radiation hard design. 
These performance studies have been repeated for the ATLASPix1\_Simple prototype and first results were reported in~\cite{Kiehn:2019wpe}.
Results from a more comprehensive study are summarized in the following.

Samples of ATLASPix1\_Simple  were irradiated with neutrons at the TRIGA Mark~II research reactor  in Ljubljana \cite{jsi} and with 17\,MeV protons at the Bern Cyclotron Laboratory \cite{bern}. Single hit efficiencies were measured in testbeams for various irradiated sensors with fluences up to several $10^{15}$\,n$_\textrm{eq}$/cm$^2$ as function of the threshold and bias voltage.
Figure~\ref{fig:eff_irrad_neutron} shows exemplarily a threshold scan for 80\,$\Omega$cm samples irradiated with neutrons for three fluences up $2 \cdot 10^{15}$\,n$_\textrm{eq}$/cm$^2$. All efficiency curves are very similar and show no significant loss of efficiency for higher fluences.
However, differences are observed in the noise levels which increase with fluence; for fluences $\geq 10^{15}$\,n$_\textrm{eq}$/cm$^2$, comparator thresholds of $\lesssim$50\,mV can only be reached by masking noisy pixels.

In order to compare the maximum achievable sensor efficiencies at similar noise levels, 
the most noisy pixels are excluded (masked) until a reasonable noise level is reached, i.e. when the average pixel noise is below 40\,Hz,
corresponding to a noise rate of $10^{-6}$ per pixel and LHC bunch crossing.
All sensors were untuned, i.e. a common comparator threshold above noise was applied to all pixels.
The resulting  maximum efficiencies after neutron and proton irradiation are shown in table~\ref{table:irrad_eff}.
All sensors show a high efficiency above 98\% for fluences of up to $2 \cdot 10^{15}$ n$_\textrm{eq}$/cm$^2$ at high bias voltages of up to 80-85\,V.
The total ionising doses (TID) of the proton-irradiated sensors are here 10\,MRad (48\,MRad) for fluences of $1 \cdot 10^{14}$ ($5 \cdot 10^{14}$) n$_\textrm{eq}$/cm$^2$.
Note that the sensor irradiated to a fluence of $1 \cdot 10^{14}$  n$_\textrm{eq}$/cm$^2$ has been biased during irradiation and was working perfectly.

\section{Summary and Outlook}
\begin{figure}[b]
\centering
\begin{subfigure}{0.46\textwidth}
\centering
\includegraphics[width=0.6\textwidth]{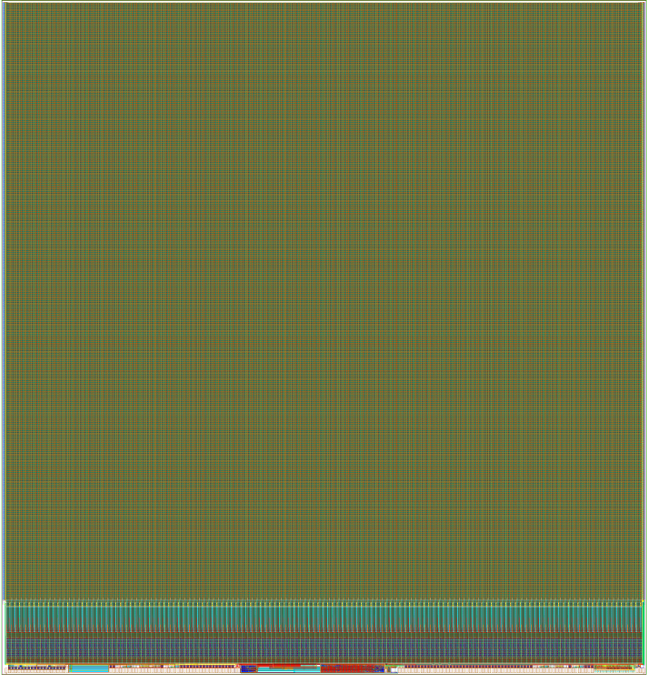}
\subcaption{\textsl{The ATLASPix3 size is about 1.9 $\times$ 2.1\,cm$^2$. The periphery is located at the lower edge, containing the readout logic for the triggered mode and test mode (continuous RO), the state machine, VCO and PLL, and the serial driver.}}
\label{fig:atlaspix3_cad}
\end{subfigure}
\qquad
\begin{subfigure}{0.48\textwidth}
\includegraphics[width=1.00\textwidth]{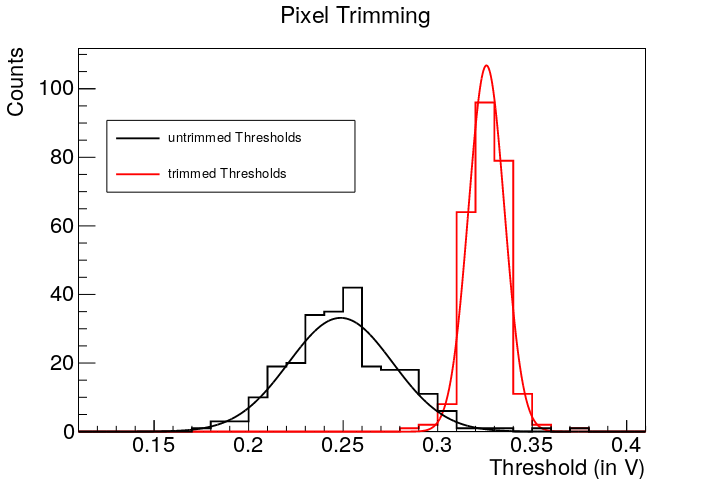}
\subcaption{\textsl{Threshold distribution of injected signals (equal to $^{55}$Fe) before and after trimming using the in-pixel 3-bit tune-DACs. The dispersion after trimming is 9.5\,mV, corresponding to about 50 electrons.}}
\label{fig:atlaspix3_meas}
\end{subfigure}
\caption{\textsl{Left: CAD drawing of ATLASPix3; Right: ATLASPix3 first measurements.}}
\label{fig:atlaspix3}
\end{figure}

Several HV-MAPS prototypes for different experimental applications and with different readout architectures have been developed  in the last years.
This talk focuses on two developments of fully monolithic pixel sensors: MuPix for the Mu3e experiment and ATLASPix for the outer pixel layer of the new ATLAS inner tracker.
Results from characterisation studies are presented for the MuPix8 and ATLASPix1\_Simple prototypes which have been produced in the ams AH18 process with different substrate resistivities (20-1000\,$\Omega$cm) and with different sensor thicknesses (50-725\,$\mu$m).
All prototypes show an excellent performance concerning efficiency and noise,  even after irradiation with protons and neutrons with fluences of up to $2 \cdot 10^{15}$ n$_\textrm{eq}$/cm$^2$. Time resolutions of about 6\,ns are measured for MuPix8 and ATLASPix1\_Simple after timewalk correction.
Correcting for binning effects from the on-chip time stamping a chip-internal time resolution of $\sim$3.7\,ns is derived for ATLASPix1\_Simple and ATLASPix1\_Simple\_Iso.

The very good time resolutions should be confirmed with the newest sensor, ATLASPix3, see figure~\ref{fig:atlaspix3_cad}.
It is designed to be functionally compatible to the RD53 readout chip~\cite{ref:rd53} and was produced in the 180~nm process by TSI~\cite{ref:tsi}.
ATLASPix3 is the largest HV-MAPS sensor produced so far, see table~\ref{tab:HVMAPS_Overview}, and was delivered in August 2019.
ATLASPix3 is fully operational and a first result is shown in figure~\ref{fig:atlaspix3_meas}.
If the ongoing measurements follow its predecessor ATLASPix1 in terms of efficiency, noise, time resolution and radiation hardness, then we would expect ATLASPix3 to fully meet the requirements of a sensor for the outer pixel tracking layer of the new ATLAS ITk.
The large size of the pixel matrix of almost 4\,cm$^2$ could also be suitable for detector modules in applications outside of ATLAS.

For March 2020 the delivery of a new MuPix10 sensor is expected, for specs see table~\ref{tab:HVMAPS_Overview}.
This sensor is also produced by TSI and has an active area  of about $2 \times 2\,$cm$^2$, close to the maximum possible reticle size.
MuPix10 has several improvements compared to previous MuPix designs (e.g. reduced cross talk) and implements  new features like voltage regulators.
It is the first protoype, which is fully compatible for the installation in the  Mu3e Pixel Tracker.
MuPix10  will be used for the construction of Mu3e pixel module prototypes which are planned to be tested at PSI in 2020.

\section*{Acknowledgments}
We thank Prof. Vladimir Cindro and Dr. Igor Mandi\'c from JSI, Ljubljana, for their help in the neutron irradiation
at the TRIGA Mark~II research reactor, supported by the H2020 project
AIDA-2020, GA no. 654168. We also thank to DESY and PSI for providing
the test beam facilities.

\end{document}